\documentclass[useAMS,aasms,astrobib,usenatbib]{mn2e}
\usepackage{epsfig,amssymb,graphicx,rotating,color,etoolbox}

\usepackage[colorlinks,linkcolor=blue,citecolor=blue,urlcolor=blue]{hyperref}        
\usepackage[all]{hypcap}

\makeatletter

\patchcmd{\NAT@citex}
  {\@citea\NAT@hyper@{%
     \NAT@nmfmt{\NAT@nm}%
     \hyper@natlinkbreak{\NAT@aysep\NAT@spacechar}{\@citeb\@extra@b@citeb}%
     \NAT@date}}
  {\@citea\NAT@nmfmt{\NAT@nm}%
   \NAT@aysep\NAT@spacechar\NAT@hyper@{\NAT@date}}{}{}

\patchcmd{\NAT@citex}
  {\@citea\NAT@hyper@{%
     \NAT@nmfmt{\NAT@nm}%
     \hyper@natlinkbreak{\NAT@spacechar\NAT@@open\if*#1*\else#1\NAT@spacechar\fi}%
       {\@citeb\@extra@b@citeb}%
     \NAT@date}}
  {\@citea\NAT@nmfmt{\NAT@nm}%
   \NAT@spacechar\NAT@@open\if*#1*\else#1\NAT@spacechar\fi\NAT@hyper@{\NAT@date}}
  {}{}

\makeatother

\usepackage{txfonts}

\usepackage[T1]{fontenc}
\usepackage{aecompl}

\usepackage{txfonts}
\title[Circumbinary disc precession]{Precession and accretion in circumbinary discs:\\The case of HD 104237}
\author[A. C. Dunhill, J. Cuadra \& C. Dougados]{A. C. Dunhill\textcolor{blue}{$^{1,2,3}$\thanks{E-mail:
\href{mailto:adunhill@astro.puc.cl}{adunhill@astro.puc.cl}}}, J. Cuadra\textcolor{blue}{$^{1,2}$} and C. Dougados\textcolor{blue}{$^{4,5}$}\\
$^{1}$Instituto de Astrof\'isica, Pontificia Universidad Cat\'olica de Chile, Vicu\~na Mackenna 4860, 7820436 Macul, Santiago, Chile\\
$^{2}$Millenium Nucleus `Protoplanetary Disks in ALMA Early Science', Santiago, Chile\\
$^{3}$Department of Physics \& Astronomy, University of Leicester, Leicester, LE1 7RH\\
$^{4}$Universit\'e Joseph Fourier / CNRS-INSU, IPAG / UMR 5274, F-38041 Grenoble, France\\
$^{5}$UMI-FCA, CNRS/INSU, France (UMI 3386), and Universidad de Chile, 1058 Santiago, Chile}
\begin{document}
\voffset=-0.5in

\date{Accepted 2015 February 9. Received 2015 February 9; in original form 2014 November 3}

\pagerange{\pageref{firstpage}--\pageref{lastpage}} \pubyear{2014}

\maketitle

\label{firstpage}

\begin{abstract}
We present the results of smoothed particle hydrodynamics (SPH) simulations of the disc around the young, eccentric stellar binary HD 104237. We find that the binary clears out a large cavity in the disc, driving a significant eccentricity at the cavity edge. This then precesses around the binary at a rate of $\dot{\varpi} = 0.48^{\circ}T_{\mathrm{b}}^{-1}$, which for HD 104237 corresponds to a precession period of 40 years. We find that the accretion pattern into the cavity and onto the binary changes with this precession, resulting in a periodic accretion variability driven purely by the physical parameters of the binary and its orbit. For each star we find that this results in order of magnitude changes in the accretion rate. We also find that the accretion variability allows the primary to accrete gas at a higher rate than the secondary for approximately half of each precession period. Using a large number of 3-body integrations of test particles orbiting different binaries, we find good agreement between the precession rate of a test particle and our SPH disc precession. These rates also agree very well with the precession rates predicted by the analytic theory of \citet{leunglee13}, showing that their prescription can be accurately used to predict long-term accretion variability timescales for eccentric binaries accreting from a disc. We discuss the implications of our result, and suggest that this process provides a viable way of preserving unequal mass ratios in accreting eccentric binaries in both the stellar and supermassive black hole regimes.
\end{abstract}

\begin{keywords}
accretion, accretion discs -- hydrodynamics -- circumstellar matter -- binaries: close -- stars: individual: HD 104237 -- stars: pre-main-sequence
\end{keywords}

\section{Introduction}\label{sec:intro}

It is widely recognised that most sun-like stars form in binary systems \citep*[e.g.][]{duquennoymayor91,raghavanetal10,jansonetal12,derosaetal14} and the binary fraction increases towards earlier spectral type. It is also known that binarity increases towards younger ages in pre-main sequence objects \citep[e.g.][]{reipurthetal14}. It should therefore be unsurprising that binaries are common in Herbig Ae/Be systems \citep[$68 \pm 11$ per cent of the full Herbig Ae/Be population, with $35$ per cent being spectroscopic binaries;][]{corporonlagrange99,bainesetal06}, where intermediate mass ($2 \lesssim M_{\star} \lesssim 8\,\mathrm{M}_{\sun}$) stars at young ages are accompanied by gaseous circumstellar (or as it may be, circumbinary) discs.

It is also well known that circumbinary discs can affect the dynamical evolution of the binary through resonant interactions, acting as an angular momentum reservoir and changing the orbital elements of the binary \citep{artymowiczetal91,artymowiczlubow94}. The direction of influence is not one-way however, and the potential of the binary dictates the form of the inner gap or cavity formed in the disc \citep{artymowiczlubow96} and how gas is able to accrete through it \citep{artymowiczlubow94}.

The importance of how binaries interact with the inner part of the disc is beginning to gain widespread recognition. The accretion process inside the central cavity is vital for understanding how both stars and their discs evolve and there are an increasing number of observations seeking to characterise this process \citep[at first from SED modelling and now from resolved mm/sub-mm images, e.g.][]{jensenmathieu97,andrewsetal11,harrisetal12}. Very recently, \citet{dutreyetal14} found evidence of ongoing planet formation in the disc around the young binary GG Tau A, identifying accretion streams flowing from the outer disc and feeding a subdisc around one star. Understanding how gas flows from the outer circumbinary disc into the the central cavity is therefore crucial to understanding planet formation processes in these systems. In particular, close binaries (common in Herbig systems) offer the opportunity to study this process as a function of the binary's orbital phase as the dynamical times in these systems are short.

Due to the non-Keplerian potential it creates, even a circular or equal-mass binary can create an eccentric shape to the disc cavity \citep[e.g.][]{macfadyenmilosavljevic08}, and this is indeed observed in widely-studied objects such as AK Sco \citep{gomezdecastroetal13}. An unequal-mass binary can also cause the cavity to become strongly decentred from the binary centre of mass as it takes on this eccentric shape. In addition, the non-Keplerian potential of the binary can force the disc's orbital elements to change in the same way that a circumbinary planet's elements osculate \citep{leunglee13} .

For solar-type and intermediate mass binary stars on the main sequence, the mass ratio $q_{\mathrm{b}}$ is roughly flat for most stellar types and companion masses \citep[e.g.][]{raghavanetal10,jansonetal12}, but with a strong bias towards equal masses at small binary separations \citep{derosaetal14}. However, \citet{kouwenhovenetal05} found that for a sample of young A-stars, the mass ratio is biased towards unequal masses, indicating that accretion at young ages plays an important role in changing the primordial mass ratio into the one observed in older systems. This is supported by simulations that find accretion occurs primarily onto the secondary \citep[e.g.][]{artymowicz83,bate00,devalborroetal11,bate14}, because its Roche lobe is further from the barycentre of the system and the relative velocity between the accreting gas and the secondary is lower. This allows the secondary to accrete more gas, pushing the mass ratio towards unity \citep*{cuadraetal09,roedigetal11,clarke12,youngetal14}. However, moving towards higher eccentricity can change this, and \citet{roedigetal11} found that for high eccentricity, the binary components can accrete at comparable rates with a mass ratio of $q_{\mathrm{b}} = 1/3$. This should affect the mass ratio of the final binary, and \citet{halbwachsetal03} indeed found that for stellar `twins', there is a small preference towards lower eccentricities, indicating that more eccentric binaries have less equal mass ratios. However, this has not been studied in detail to date.

Simulations studying accretion in eccentric binary systems have shed a little light on this process. \citet{guntherkley02} found that the accretion rate was dependent upon orbital phase for a number of eccentric systems .More recently, \citet{shietal12} found that an eccentric binary drives a disc precession in high-resolution magneto-hydrodynamic (MHD) simulations, but did not run the simulation for more than a fraction of the precession timescale. \citet{gomezdecastroetal13} also found an eccentric and decentred disc cavity in simulations of GG Tau A, but again the timescale of their simulations was short.

The modelling work presented in this paper has been motivated by the Herbig Ae system HD104237 (DX Cha). HD104237 is a nearby \citep[$d = 116 \pm 7$ pc;][]{perrymanetal97} spectroscopic binary (SB2) with a semimajor axis $a_{\mathrm{b}} = 0.22$ au, eccentricity $e=0.6$ and orbital period $T_{\mathrm{b}}  = 20$ days \citep{bohmetal04,garciaetal13}. The total stellar mass is $3.6$ M$_{\sun}$ and the mass ratio inferred from the spectral type of the primary is $q_{\mathrm{b}} = 0.64$. The system is inclined by $i \lesssim 20^{\circ}$ to the plane of the sky \citep{gradyetal04,bohmetal04,garciaetal13}.

Because of its near face-on geometry, proximity and brightness, this system is currently a unique and prime target for detailed study of the accretion process in a close binary system with long baseline interferometry. \textit{AMBER}/\textit{VLTI} observations show that most of the K band continuum flux arises from a ring of radius $R = 0.5$ au \citep{tatullietal07}, in agreement with the location of the dust sublimation radius in the circumbinary disc \citep{garciaetal13}. However significant unresolved emission is present within this radius, in excess of the expected contributions from the photospheres. The stars in this system probably do not hold individual circumstellar discs due to outer tidal truncation at closest binary approach.

This unresolved flux likely arises in compact structures (e.g. accretion streams) inside the tidally disrupted circumbinary disc \citep{garciaetal13}. The AMBER observations also show that most ($90$ per cent) of the HI Br$\gamma$ emission is unresolved and comes from the immediate vicinity of the stars. An increase in Br$\gamma$ emission is observed on each binary component at periastron passage, likely related to an accretion burst \citep{garciaetal13}, similarly to what has been observed previously. We have recently observed this system with \textit{PIONIER}/\textit{VLTI} in the H band at different orbital phases, with the aim of reconstructing the morphology of this tidally disrupted circumbinary disc and study the evolution of the accretion process onto the central binary with orbital phase. The description and analysis of these data, in particular their detailed comparison to the results of the modelling work presented here, will be presented in a forthcoming paper (Dougados et al., in prep).

To build on and extend the findings of work by previous authors \citep[e.g.][]{shietal12,gomezdecastroetal13}, we perform smoothed particle hydrodymanics (SPH) simulations of this system, with an emphasis on exploring the long-term evolution of accretion patterns in the system (over more than 1000 orbital periods). The structure of the paper is as follows. In Section \ref{sec:sims} we describe the numerical parameters of our simulations and the method used to carry them out, and in Section \ref{sec:results} we describe the results. In Section \ref{sec:apply} we apply these findings to the general case of a binary of any given eccentricity and compare the precession timescale of our SPH simulation with the predictions of \citet{leunglee13}. We discuss the implications for the study of discs around stellar binaries, in particular the role of accretion in eccentric systems, and the limitations of our methods and model in Section \ref{sec:discussion}, before presenting our conclusions in Section \ref{sec:conclusions}.

\section{Simulations}\label{sec:sims}

We have simulated the HD 104237 system using the hybrid SPH/$N$-body code \textsc{Gadget-2} \citep{springel05}, modified to closely follow the dynamics of the binary and to include a Navier-Stokes viscosity \citep*{dunhilletal13}. We model the binary components as $N$-body particles with a mass ratio $q_{\mathrm{b}} = 0.64$ (so that $M_{\mathrm{A}} = 2.2 \mathrm{\,M}_{\sun}$ and $M_{\mathrm{B}} = 1.4 \mathrm{\,M}_{\sun}$), semimajor axis $a_{\mathrm{b}} = 0.22$ au and eccentricity $e_{\mathrm{b}} = 0.6$.

We model the disc using SPH particles, with $N$ = 2 million particles for our main simulation. We paramaterize the disc viscosity using the \citet{shakurasunyaev73} $\alpha$-disc prescription, with $\alpha = 10^{-2}$. The disc initially extends between $1.5 < R < 20\,a_{\mathrm{b}}$, with a surface density profile that approximately follows $\Sigma \propto R^{-1.7}$, but is not a strict power-law\footnote{Our initial conditions were generated by azimuthally averaging the surface density profile of the final snapshot from the simulations described by \citet{cuadraetal09}.}. We impose an isothermal equation of state on the disc, so that the temperature (and thus the disc scale-height $H$) is a function only of radius from the barycentre of the binary. We choose the temperature profile so that it results in a flaring disc where $H/R \propto R^{1/4}$, normalised so that $H/R = 0.05$ at $R = 1$ au. We model the stellar accretion using a simple sink particle method, where each star is modelled as a point mass with an associated sink radius, $R_{\mathrm{sink}} = 0.03\,a_{\mathrm{b}}$. Any SPH particle found within this radius has its mass and momentum added to the star, and the SPH particle is removed from the simulations. We discuss the physicality of this implementation and its implications in Section \ref{discussion:limitations}. For calculations of the accretion rate discussed in Section \ref{sec:results} we refer to the rate at which SPH particles are `accreted' by the stars in this way. The binary components are modelled gravitationally as point masses and their orbit is free to evolve through the simulation. As the disc has a low mass this only occurs at a level below one per cent of the initial orbital parameters on the timescales of our simulation, and we therefore do not report these changes.

Our disc has a mass $M_{\mathrm{d}} = 5 \times 10^{-3} M_{\mathrm{b}} = 0.018$ $\mathrm{\,M}_{\sun}$. \citet{halesetal14} estimated the total disc mass to be $M_{\mathrm{d}} = 0.04$ $\mathrm{\,M}_{\sun}$ from their SED modelling. Their model suggests a characteristic radius of 90 au, and our simulated disc is much smaller than this, so our disc is likely to be over-massive. The smaller radial extent of our disc is due to numerical reasons, but our neglect of all but the inner regions is justified as the the outer part of the disc should not play any dynamical role in what happens in the centre. However it does affect the mass in the inner regions as the disc spreads viscously, and we therefore normalise the accretion rates discussed in Section \ref{sec:results} to account for the fact that while the surface density in our disc decreases with time, in the real disc around HD 104237 this process occurs only on time-scales much longer than we simulate. Since we are well below the self-gravitating limit the mass effectively becomes a scaling factor and does not affect the hydrodynamics. SED models of HD 104237 are consistent with a non-flaring disc \citep{fangetal13,halesetal14}, while our equation of state enforces a flared disc. The flaring is not strong, and $H/R$ only goes from 0.03 at the inner edge to 0.07 at the outer edge of our disc, so we do not expect this to affect the results of our simulations. Our model also does not include stellar winds or the jet associated with the system.

We use a set of code units such that the distance unit $R_0 = a_{\mathrm{b}}$, the mass unit $M_0 = M_{\mathrm{A}} + M_{\mathrm{B}}$ and the time unit $T_0 = T_{\mathrm{b}} / 2\pi$ where $T_{\mathrm{b}}$ is the orbital period of the binary. This choice of units sets the gravitational constant $G = 1$ internally to the code.

\begin{figure}
\includegraphics[width=\columnwidth]{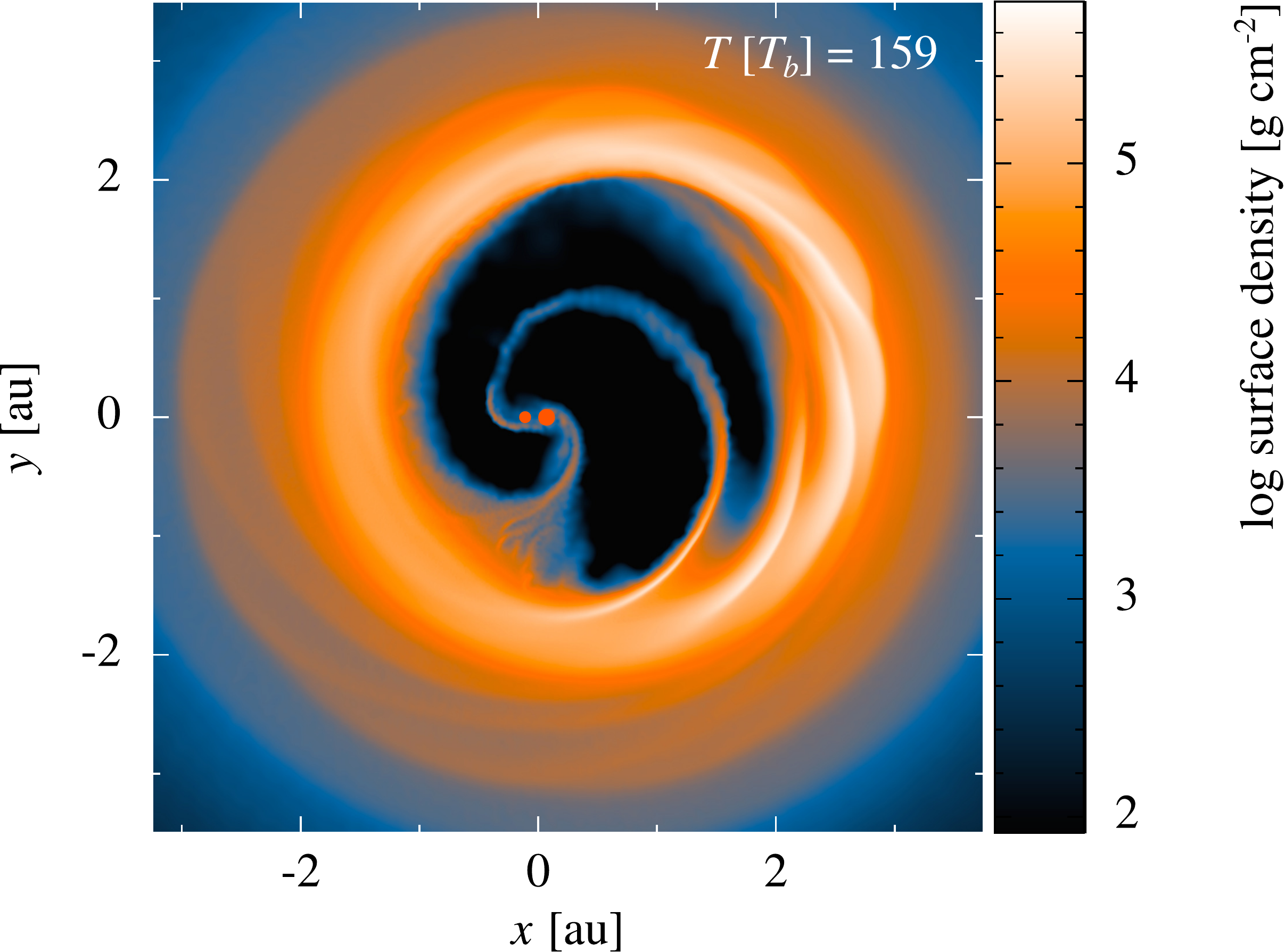}
\caption{Snapshot of our reference simulation with $N = 2$ million particles after 159 binary orbits, with the binary at pericentre. The clearly eccentric cavity shape is formed by the unequal-mass, eccentric binary `slingshotting' material across the cavity and pushing the far wall of the cavity further from the binary.}
\label{fig:1}
\end{figure}

\begin{figure*}
\includegraphics[width=\linewidth]{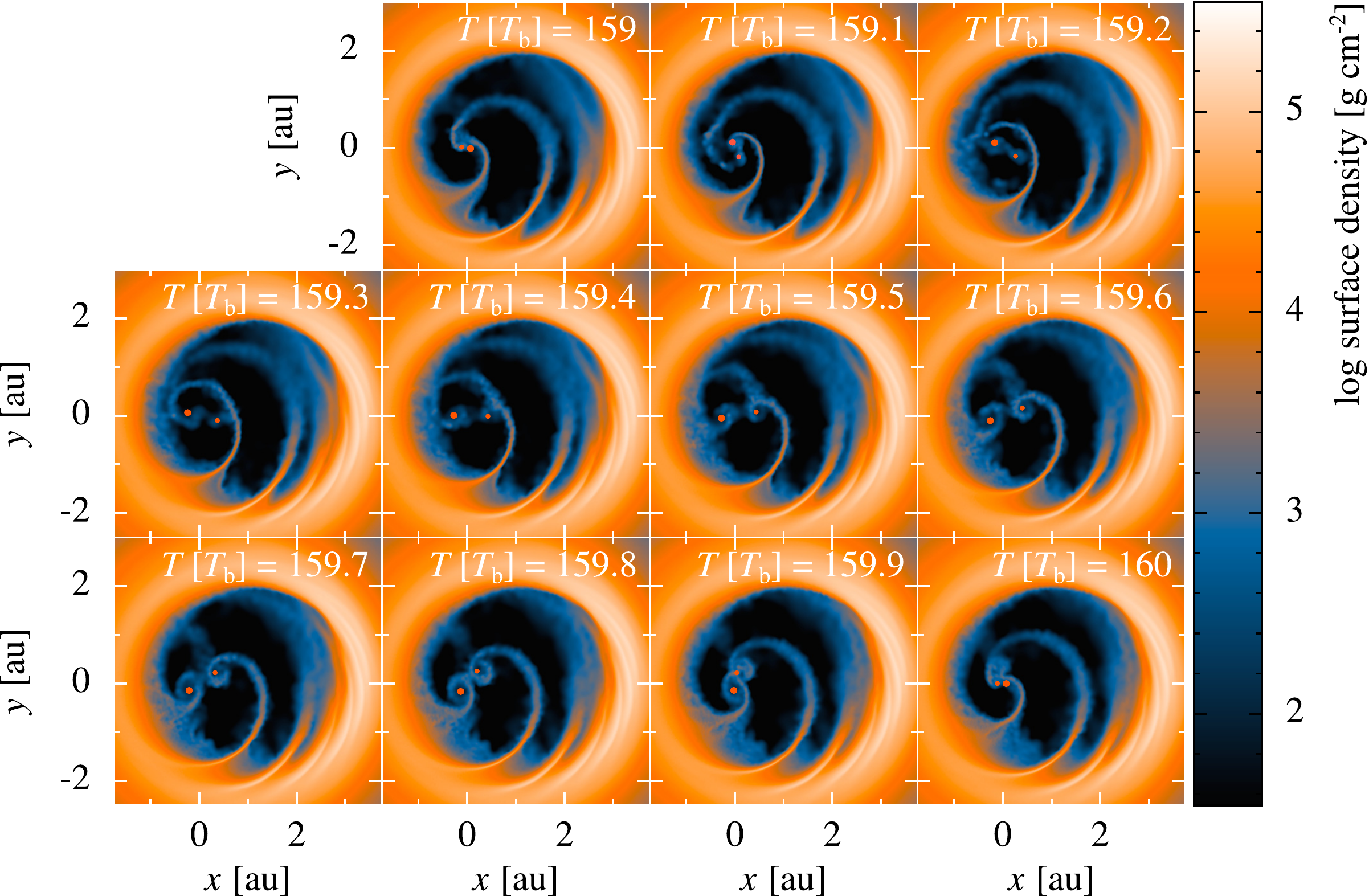}
\caption{Series of snapshots showing the accretion process as a function of the binary's orbital phase, with time $T$ in units of the binary period $T_{\mathrm{b}}$. The binary is at pericentre in the uppermost left panel (the same snapshot as in Figure \ref{fig:1}), and at apocentre in the panel third from left in the middle row. The primary (larger circle) can be seen to drag disc material in from the nearest cavity edge at apocentre passage, which is then thrown across the cavity due to the binary's eccentric orbit. This stream is then partially dragged back towards the binary by the secondary when it next reaches apocentre, and this material is accreted by the binary components near pericentre. As the cavity precesses with respect to the binary orbit, which binary component plays the role of creating the accretion stream and which drags it back to the binary will switch.}
\label{fig:2}
\end{figure*}

After the initial conditions settle, the action of the binary drives an eccentric inner cavity in the disc. Material drawn in as a tidal tail by the primary is then kicked across the cavity, and pushes the far cavity wall away from the binary. A snapshot showing this process is shown in Figure \ref{fig:1}, and we show how this evolves as a function of orbital phase in Figure \ref{fig:2}. The eccentric cavity shape is then key to the evolution of the system in our simulations.

The inner region of the disc then begins to precess with respect to the orbit of the binary, as shown in Figure \ref{fig:3}. This precession is due to the highly non-Keplerian nature of the potential close to an unequal-mass eccentric binary such as HD 104237, and is a natural consequence of the fact that orbits in a non-Keplerian potential are not closed ellipses. To date, this effect has only been studied for discs around circular binaries in the context of supermassive black hole (SMBH) binaries \citep*{macfadyenmilosavljevic08,shietal12,dorazioetal13,farrisetal14}. We find that the precession is uniform between $4\lesssim R \lesssim 6\,a_{\mathrm{b}}$. We ran the simulation for 1500 binary orbits before arbitrarily halting it.

\begin{figure}
\includegraphics[width=\columnwidth]{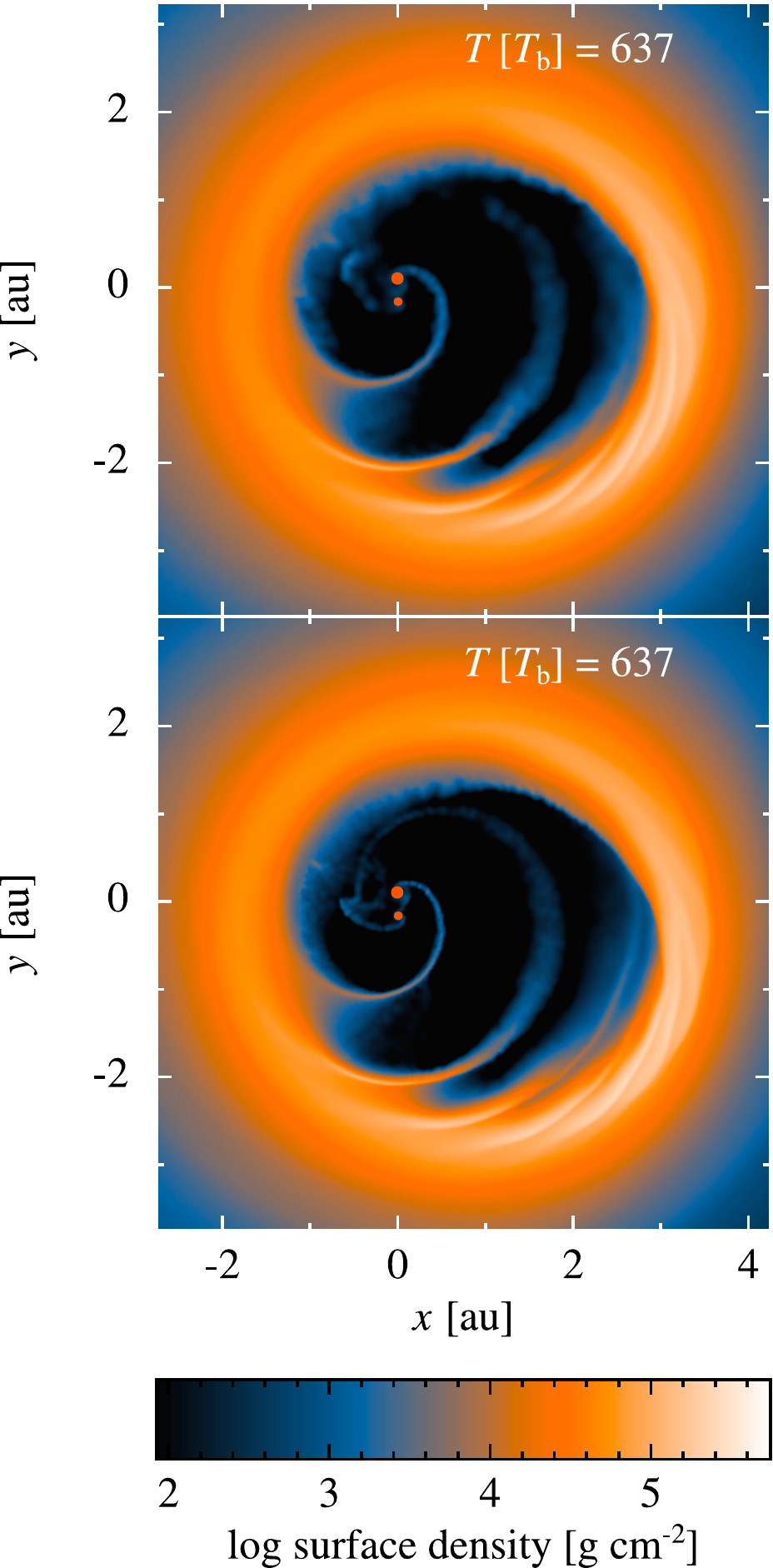}
\caption{Comparison between the original (top panel) and higher-resolution run (bottom panel) at the end of the latter run. The accretion streams within the cavity are noticeably better-resolved in the higher-resolution version, and the cavity shape is slightly more eccentric. We compare cavity precession rates and the measured accretion rates between the two runs in Figures \ref{fig:5} and \ref{fig:7}.}
\label{fig:4}
\end{figure}

As our simulation is only of moderate resolution (especially inside the cavity), we wanted to conclusively rule out the possibility that this precession rate was affected by any anomalous resolution or viscosity issues in the SPH. We therefore took a snapshot after approximately $T = 450\,T_{\mathrm{b}}$ and resampled it, replacing each SPH particle with 4 new particles at positions randomly sampled from the smoothing kernel and the parent particle's smoothing length. We then ran this higher resolution version of the simulation for a further 180 binary orbits. Although the up-samping introduces noise into the new simulation at the very start, this is soon dissipated and the simulation follows the same course as the original. A comparison between the original and up-sampled simulations at the time we halted the higher resolution run can be seen in Figure \ref{fig:4}. The only significant difference between the two simulations is that in the higher-resolution version the accretion rate is lower as we are better able to resolve the individual accretion streams.

\section{Results}\label{sec:results}

\begin{figure}
\includegraphics[width=\columnwidth]{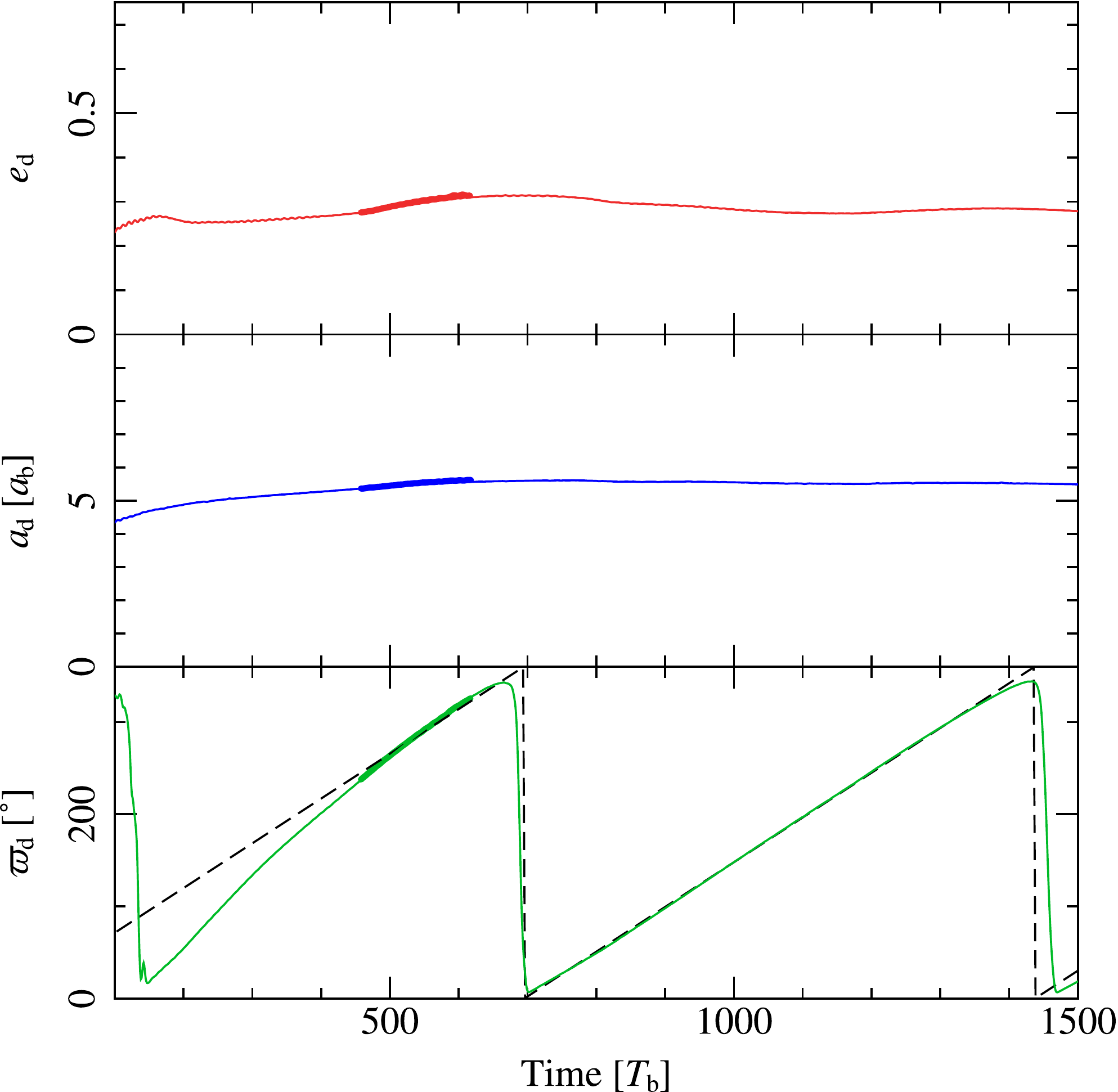}
\caption{Evolution of the disc eccentricity ($e_{\mathrm{d}}$, upper panel), semimajor axis ($a_{\mathrm{d}}$, middle panel) and argument of periaps ($\varpi_{\mathrm{d}}$, lower panel). Thin solid lines are for our main simulation and bold lines are for the shorter higher-resolution version. These values are averaged between $4.5 < R < 5.5\,a_{\mathrm{d}}$, where the disc precesses uniformly. In the lower panel, we also plot a best fit to the precession, where $\dot{\varpi} = 0.48\,^{\circ}\,T_{\mathrm{b}}^{-1}$.}
\label{fig:5}
\end{figure}

In order to characterise the evolution of the inner region of the disc, we calculate the eccentricity $e_{\mathrm{d}}$, semimajor axis $a_{\mathrm{d}}$ and argument of periapse $\varpi_{\mathrm{d}}$ in this region of uniform precession using
\begin{equation}
\xi_{\mathrm{d}}(R) = \frac{\int_0^{2\pi} \Sigma(R,\phi) \xi d\phi}{\int_0^{2\pi} \Sigma(R,\phi)\,d\phi},
\label{eq:element}
\end{equation}
where $\xi$ is any of the orbital elements and $\Sigma(R,\phi)$ is the disc surface density. We then average over the region of uniform precession, and plot the results in Figure \ref{fig:5}. We also plot the same values for the high-resolution run, and find excellent agreement especially for the disc precession. The precession rate is initially faster than its final value before decelerating and converging on $\dot{\varpi} = 0.48\,^{\circ}T_{\mathrm{b}}^{-1}$. In physical units this corresponds to a precession period for the disc around HD 104237 of $T_{\mathrm{prec}} \sim 40$ years.

\begin{figure}
\includegraphics[width=\columnwidth]{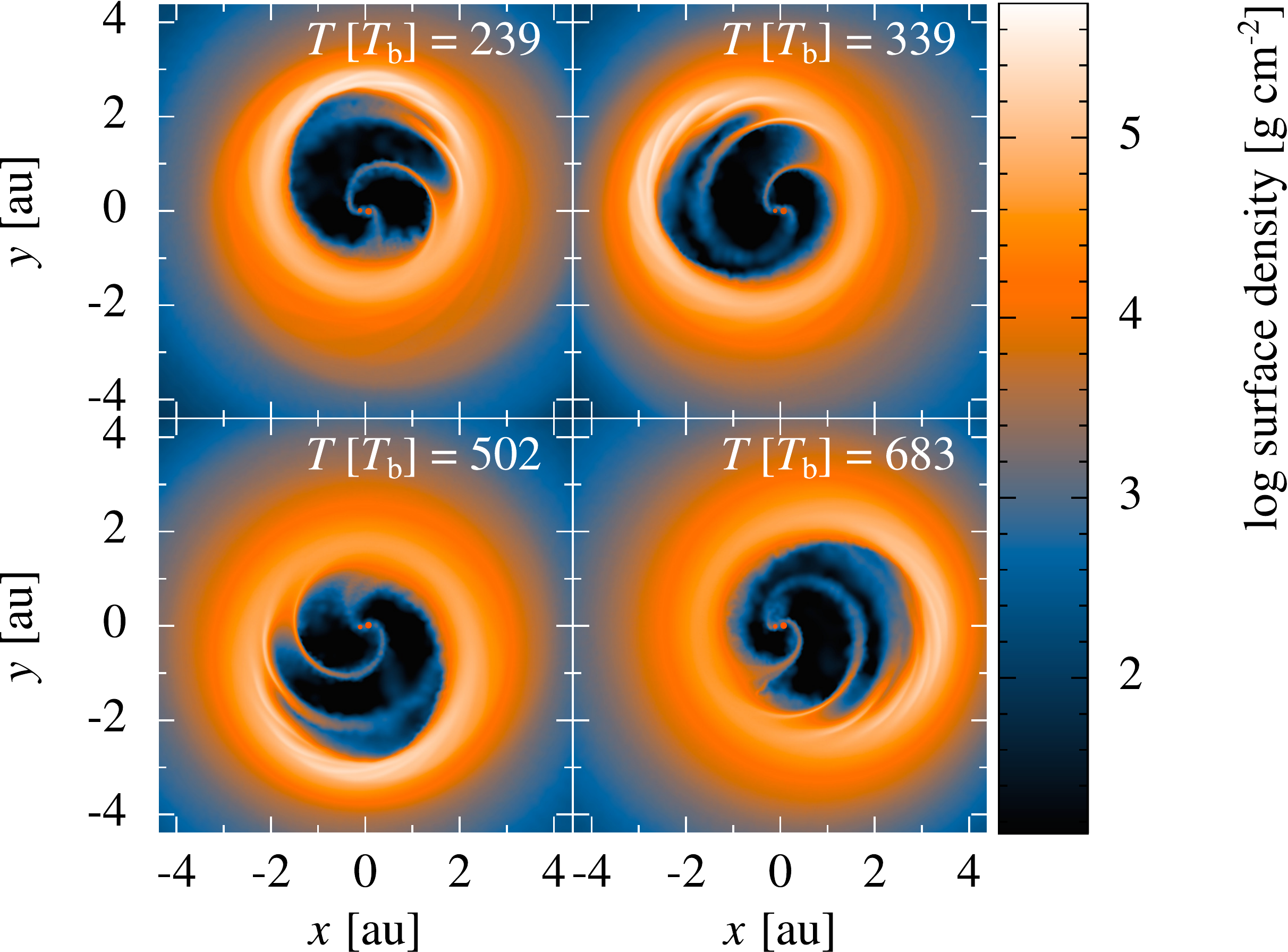}
\caption{Series of snapshots showing the precession of the cavity. Times are in units of the binary orbit and the binary is at pericentre in each snapshot. At these early stages in the simulation the precession rate is changing so the time interval between the panels is not uniform (see Figure \ref{fig:5}).}
\label{fig:3}
\end{figure}

\begin{figure}
\includegraphics[width=\columnwidth]{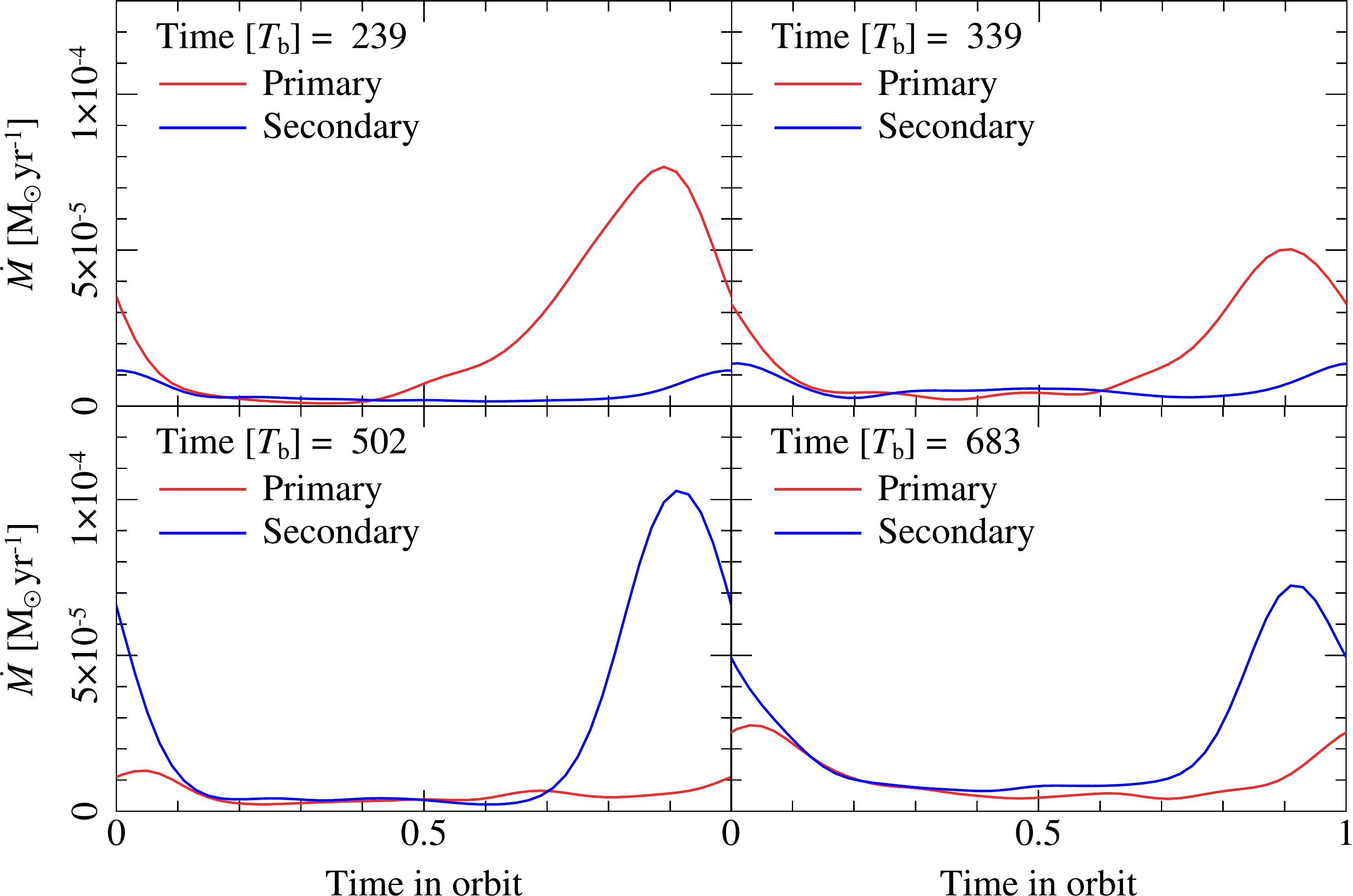}
\caption{Accretion rates as a function of orbital phase at 4 different times during the simulation. Times 0 and 1 correspond to the binary at pericentre and 0.5 corresponds to the binary at apocentre. The accretion rates are averaged over 10 binary orbits, and smoothed using a moving-average algorithm. It is noteworthy that the binary component with the highest accretion rate changes as the simulation evolves, as this is dictated by the precession of the disc cavity. Also of note is that the highest accretion peak occurs just before pericentre, and the lower accretion peak occurs at apocentre. This is due to how the accretion proceeds within the cavity (see Figure \ref{fig:2}), but is not necessarily correct due to our neglect of proper thermodynamic treatment (see Section \ref{sec:discussion}), }
\label{fig:6}
\end{figure}

How the accretion proceeds inside the cavity is strongly dependant upon the position angle of the cavity. This results in a periodicity in the accretion signature in our simulation that is wholly tied to the precession of the disc. In Figure \ref{fig:6} we show the accretion as a function of orbital phase at the same 4 points in the simulation as the panels of Figure \ref{fig:3}, showing that the peak accretion rate per orbit, and which component it is on to, changes with the position angle of the cavity. To generate this figure we average over 10 orbits either side of the snapshots shown in Figure \ref{fig:3} in order to reduce noise and highlight this short-term trend. 

\begin{figure}
\includegraphics[width=\columnwidth]{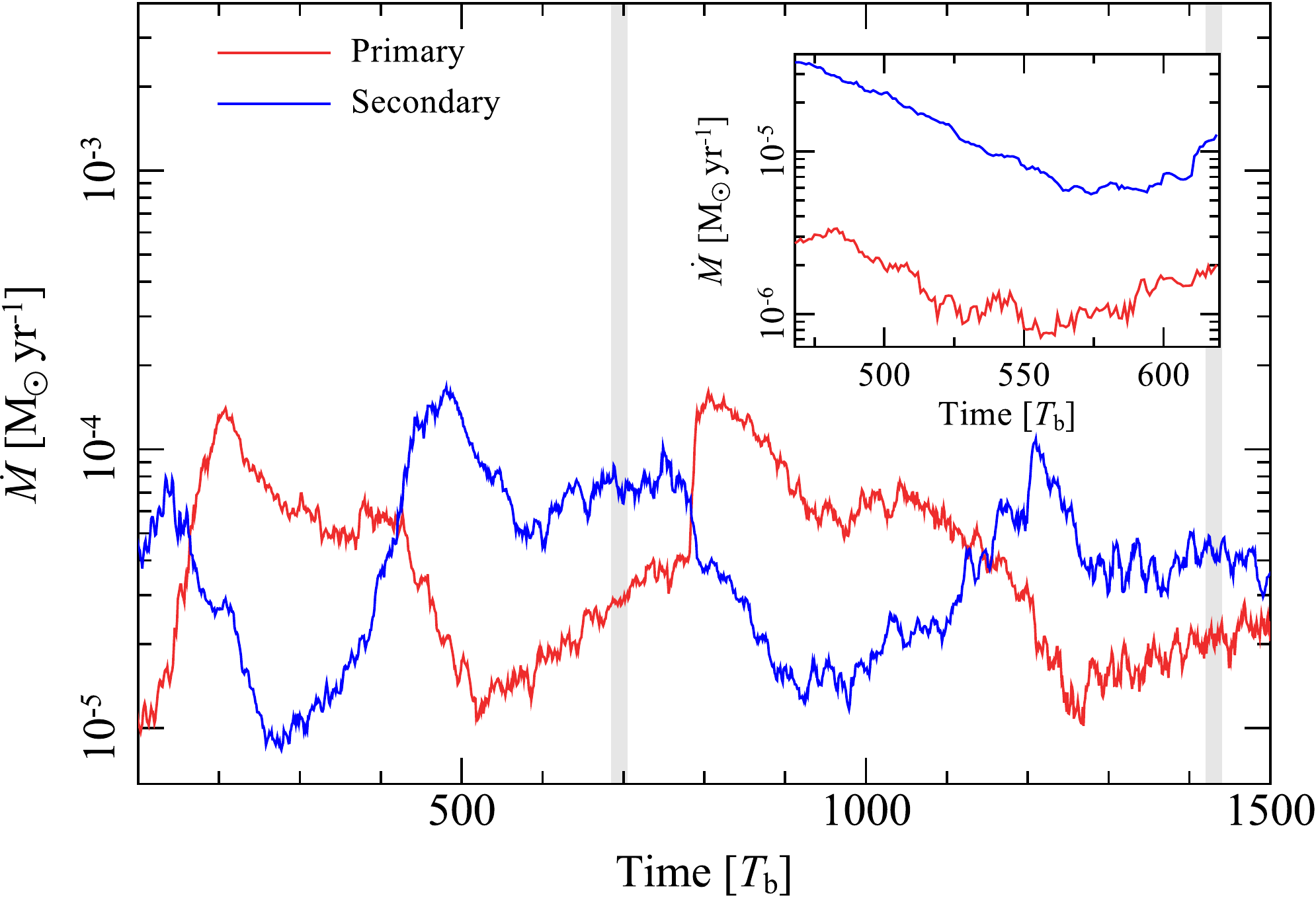}
\caption{Measured accretion rates in our main simulation, with the accretion rates for the up-sampled higher resolution run inset. The periodic change in the accretion rate is dictated by the changing position angle of the cavity, which can be easily seen by comparing this with the lower panel of Figure \ref{fig:5}. The accretion rates here are a factor of $10^3-10^4$ higher than that measured for the real system \citep[$\dot{M} \sim 10^{-8}$ M$_{\sun}$ yr$^{-1}$;][]{gradyetal04}. This is primarily due to the low resolution of our runs (the inset panel clearly shows that the increased resolution run has accretion rates that are lower by an order of magnitude) and the higher disc mass in our simulation. The higher resolution runs, in which the accretion is better (but not fully) resolved show good agreement with the general pattern of the primary run, with both primary and secondary rates turning over at $T \sim 550\,T_{\mathrm{b}}$ as the cavity precesses. Grey bars indicate the times when the binary and disc cavity position angles coincide (i.e. $\varpi_{\mathrm{b}} = \varpi_{\mathrm{d}}$).}
\label{fig:7}
\end{figure}

Figure \ref{fig:7} shows how these peak accretion rates evolve through the simulation. As we neglect the outer regions in our simulation, the surface density of the disc decreases over time due to viscous spreading. In the real HD 104237 disc, the outer disc acts as a mass reservoir so this does not occur. To account for this, we normalise the accretion rate to the remaining disc mass within $R \leq 5\,a_{\mathrm{b}}$. We include a comparison with the higher-resolution resample as an inset panel, showing how increasing the resolution changes the absolute accretion rate but that the trend remains unchanged.

We see from these three figures that there are two strong periodicities obvious in the accretion rates we find in the simulation. Firstly, on a per-orbit time-scale as shown in Figure \ref{fig:6}, the binary component which sees the highest accretion rate peaks just before the binary reaches pericentre, with the lesser accretor peaking at pericentre. This is due to the fact that the binary accretes by pulling material back in from the accretion stream that has been thrown out across the cavity, and this reaches the binary just before pericentre and is accreted, with a little remaining gas then falling onto the other component just as the binary reaches pericentre (see Figure \ref{fig:2}).

On longer time-scales, and in a pattern dictated entirely by the non-Keplerian potential of the binary, we see accretion variability as the cavity precesses and the process described above is modulated by the changing amounts of gas available in the accretion streams and switching over of the binary component that creates the streams (see Figure \ref{fig:3}).
The time-scale for this process to change is half the precession period of the disc, corresponding to approximately 20 years for the real HD 104237 disc\footnote{An animation showing the accretion process and the effect of the cavity precession upon it can be found online at \href{http://www.acdunhill.com/hd104237}{www.acdunhill.com/hd104237}.}.

Taking the precession-period averaged accretion rates $\langle \dot{M}_{\mathrm{A}} \rangle / \langle \dot{M}_{\mathrm{B}} \rangle$ where $\dot{M}_{\mathrm{A}}$ and $\dot{M}_{\mathrm{B}}$ are the accretion rates onto the primary and secondary respectively, we find values between $0.9$ and $1.3$. This means that this precession mechanism allows the primary to accrete at the same rate as, and at times at an even higher rate than, the secondary. In purely physical terms then, this provides a viable mechanism for allowing an eccentric binary to accrete matter while maintaining an unequal mass ratio, something that has proven problematic for simulations of binary formation \citep[e.g.][]{clarke12,youngetal14}.

We believe that the first variability, on single-orbit time-scales, is indeed observed in the real system. As reported by \citet{garciaetal13}, the equivalent width of the Br$\gamma$ line increases around pericentre passage, and this is interpreted as evidence of enhanced accretion at this phase of the orbit. The observability of the second periodicity depends upon how the accretion luminosity and physical accretion rate are related, which is far beyond the physics included in our SPH simulation. However, it is instructive to see at what level we might expect to see the accretion to vary on these 20 year time-scales, so we use the most basic possible model for accretion luminosity -- gravitational energy lost is converted into luminosity -- to make very rough estimates. Although this model is known to be incorrect in many ways \citep[e.g.][]{darioetal14}, we do not necessarily trust all aspects of the accretion process that we do model in the simulations, so it would be over-complicating matters to apply a model too sophisticated.

Under this assumption then we have an accretion luminosity given by
\begin{equation}
L_{\mathrm{acc}} = G\frac{\dot{M}\,M_{\star}}{R_{\star}}\left(1 - \frac{R_{\star}}{R_{in}}\right),
\label{eq:lum}
\end{equation}
where $\dot{M}$ is the accretion rate, $R_{\mathrm{in}}$ is the radius from which material finally falls onto the star -- taken to be the corotation radius and assumed to be $R_{\mathrm{in}} \sim 17\,R_{\sun}$ for both stellar components \citep[][from which paper we also take values for the stellar masses $M_{\star}$]{garciaetal13}. We take the values of the stellar radii $R_{\star}$ from \citet{fumelbohm12}. The values of $\dot{M}$ we take from Figure \ref{fig:7}, but scale them down by a factor of $10^{3}$ in order to better match the observed accretion rate in the system \citep[$\dot{M} \sim 10^{-8}$ M$_{\sun}$ yr$^{-1}$;][]{gradyetal04}.

Although it is possible that the stars have small sub-discs around them which would provide a mechanism to delay accretion onto the stars themselves, we think this unlikely and do not include it in our model. This is because the truncation and stellar radii are very similar, so any disc would have to be very small indeed \citep*{pichardoetal05,garciaetal13}. Instead we model the accretion as falling directly from the corotation radius of each star \citep{darioetal14}. As the two components of the binary cannot be distinguished at pericentre, we sum their luminosities and plot the resultant accretion luminosity as a function of time in Figure \ref{fig:8}.

We find that the variability expected even for this most optimistic case (where we assume all lost gravitational energy is radiated as observable signature) the variability in the model is lower than the uncertainty in the luminosity of the primary star of the system \citep[$L_{\mathrm{A}} = 30^{+5}_{-4}\,\mathrm{L}_{\sun}$;][]{garciaetal13}. We therefore do not expect the signature of precession-driven accretion variability to be observable over decade-long time-scales if the accuracy of these measurements does not improve dramatically. However, it is possible that this signature could be observed for eccentric short-period binaries with higher accretion rates, as the signal would be stronger in such cases. Furthermore, if the uncertainties in the observed stellar luminosities are systematic rather than intrinsic, and apply equally to measurements taken years apart, it is possible that this $\Delta L_{\mathrm{acc}} \sim 2\,\mathrm{L}_{\sun}$ variability will indeed be observable.

\section{Application}\label{sec:apply}

Our primary result is that precession of the circumbinary disc can cause long-term periodic accretion variability. In order to make this more widely applicable, we ran a large number of simple 3-body integrations of a test particle orbiting binaries of different mass ratios and semi-major axes. We varied the initial eccentricity and semimajor axis of the test particle orbit, and for each configuration we ran 100 integrations with the test particle's initial orbital phase drawn at random from a uniform distribution between 0 and $2\pi$. We explored test particle semimajor axes in the range $2.5 \leq a_{\mathrm{t}} \leq 8.5$, and eccentricities $0.1 \leq e_{\mathrm{t}} \leq 0.5$. The test particle eccentricity made no difference to the precession rates of the particles in the range we explored, so we average the results over the eccentricities. We tested six different binaries, with mass ratios $q_{\mathrm{b}} = 1/3$ and $1/2$, each with eccentricities $e_{\mathrm{b}} = 0$, $0.3$ and $0.6$. In each case we integrated the system for $T=10$,$000\,T_{\mathrm{b}}$, but stopped if scattering from a close interaction caused $e_{\mathrm{t}} \geq 0.8$. For each of the $6$ binary configurations we ran $5$,$000$ integrations, for a total of $30$,$000$ runs.

\begin{figure*}
\includegraphics[width=\linewidth]{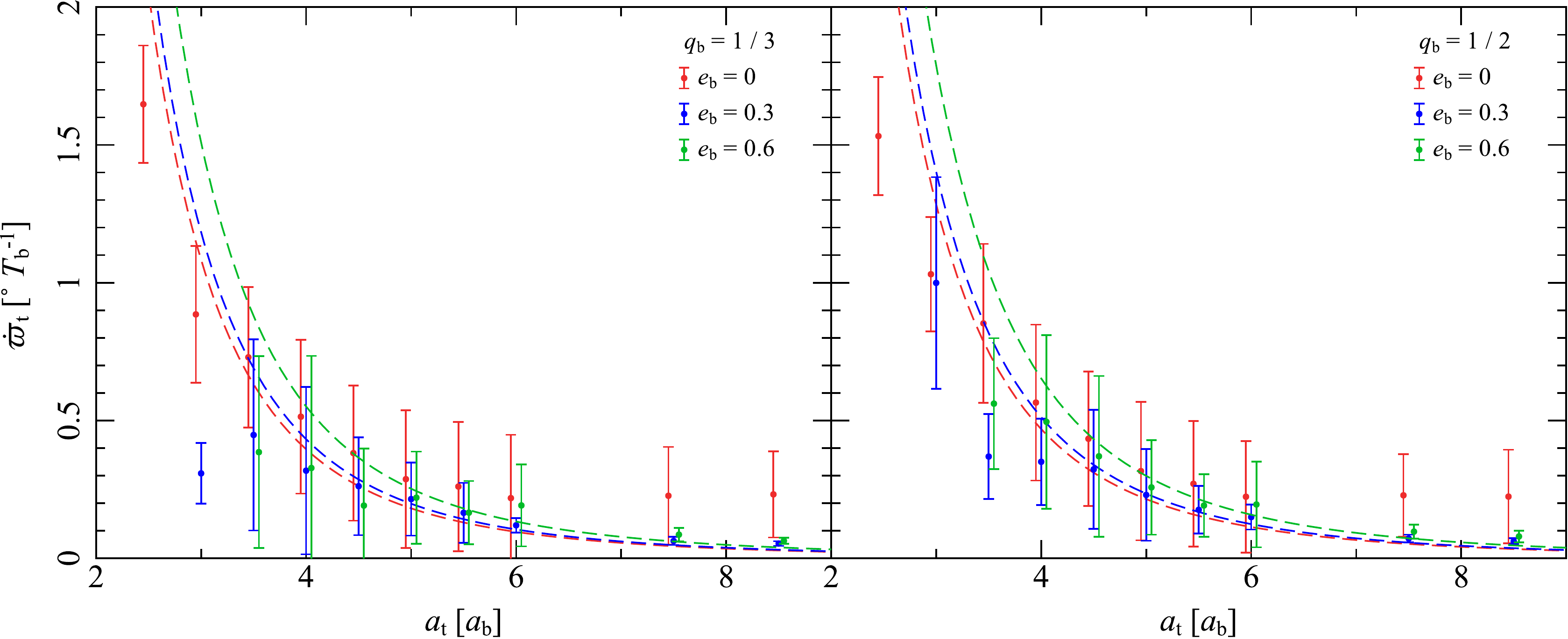}
\caption{Apsidal precession rates for test particles around binaries with mass ratio $q_{\mathrm{b}} = 1 / 3$ (left panel) and $q_{\mathrm{b}} = 1 / 2$ (right panel), and binary eccentricities $e_{\mathrm{b}} = 0$ (red points), $0.3$ (blue points) and $0.6$ (green points).  Red and green points are respectively offset slightly negatively and positively on the $x$-axis to increase visibility. Error bars are the standard deviation of normal distributions fitted to histograms of precession rates measured from the test particle runs. Dashed lines in each case give the \citet{leunglee13} prescription for the precession rates given in Equation \ref{eq:ll13}. We find excellent agreement, except for the case of circular binaries at large semimajor axis, and at small semimajor axis.}
\label{fig:9}
\end{figure*}

We then measured the change in the test particle's argument of periapse, and binned these values to create a histogram of precession rates for each configuration of binary and test particle orbit. Fitting a gaussian to this gives us a simple estimate of how the precession rate of the test particle varies with its own and the binary's orbital parameters. We plot these precession rates in Figure \ref{fig:9}, with the standard deviation of the fitted gaussian as an error bar to show the width of the precession rate distribution for each set of parameters.

\citet{leunglee13} showed that the precession rate for a test particle orbiting a binary (in the context of circumbinary planets) can be well approximated by
\begin{equation}
\dot{\varpi_{\mathrm{t}}} \simeq \frac{3}{4}\frac{M_{\mathrm{A}}\,M_{\mathrm{B}}}{(M_{\mathrm{A}} + M_{\mathrm{B}})^2}\left(\frac{a_{\mathrm{b}}}{a_{\mathrm{t}}}\right)^2\,\left(1 + \frac{e_{\mathrm{b}}^2}{2}\right)^2
\label{eq:ll13}
\end{equation}
where $\dot{\varpi_{\mathrm{t}}}$ and $a_{\mathrm{t}}$ are respectively the apsidal precession rate and semimajor axis of the test particle. The $(1+e_{\mathrm{b}}^2/2)^2$ term is a correction that makes the prescription more accurate with high binary eccentricity, and their tests show that it is approximately correct at $e_{\mathrm{b}} = 0.5$ (to within ten per cent). We plot the predicted precession rate from this prescription as dashed lines in the panels of Figure \ref{fig:9}.

We find that this prescription gives a very good approximation to our test particle precession rates, even for high eccentricity ($e_{\mathrm{b}} = 0.6$) binaries. In fact, we only find significant disagreement for two cases: at low $a_{\mathrm{t}}$, and for large $a_{\mathrm{t}}$ around circular binaries. The former is easily understood, as \citet{leunglee13} derive Equation \ref{eq:ll13} by taking only the lowest power of $a_{\mathrm{b}}/a_{\mathrm{t}}$ in an expansion. The latter disagreement is more puzzling, but as we are interested in using their theory to predict precession timescales for discs around high eccentricity binaries, finding the source of this discrepancy where $e_{\mathrm{b}} = 0$ is beyond the scope of this paper and we do not pursue it further.

\begin{figure}
\includegraphics[width=\columnwidth]{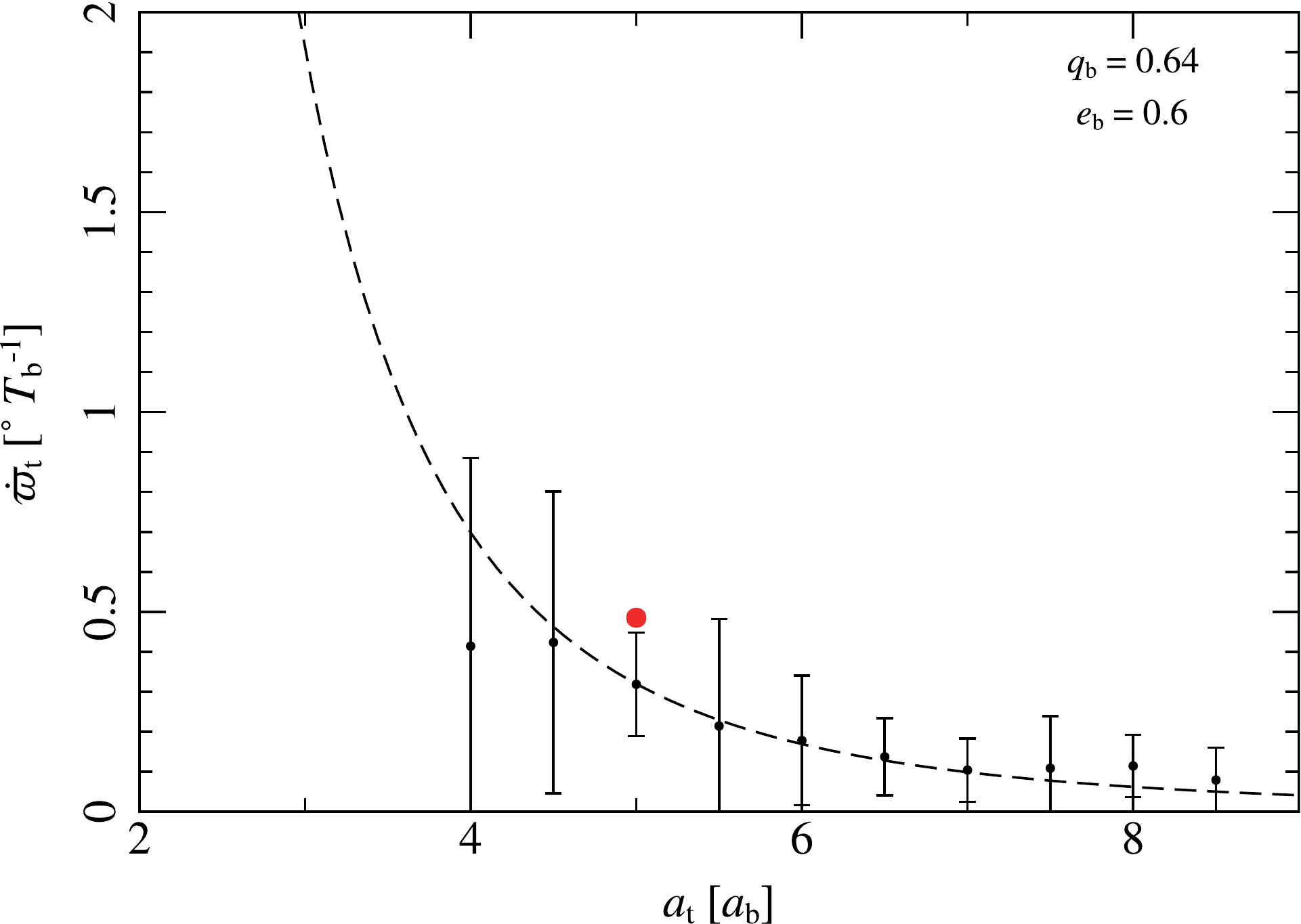}
\caption{As figure \ref{fig:9} but for a binary matching the binary in our SPH simulations. The dashed line give the \citet{leunglee13} prescription for the precession rates given in Equation \ref{eq:ll13}, and the red circle is the measured rate from our simulations (values taken from Figure \ref{fig:5}). We find good agreement between our SPH precession rate and that predicted by \citet{leunglee13}, consistent with the uncertainty from the test particle runs.}
\label{fig:10}
\end{figure}

We also repeated the test particle experiment for a binary that matches that in our SPH simulation, performing a further $8$,$000$ integrations of test particles with  $4 \leq a_{\mathrm{t}} \leq 8.5$ and $0.1 \leq e_{\mathrm{t}} \leq 0.8$, and plot the results in Figure \ref{fig:10} along with the values found for our disc shown in Figure \ref{fig:5} and the \citeauthor{leunglee13} predictions using Equation \ref{eq:ll13}. Again, we find good agreement with the prescription of \citet{leunglee13}, both for the test-particle integrations and for the SPH disc. We conclude that while resonances with the eccentric binary that act to carve the gap in the first place may have some effect on the stability of gas orbits at the cavity edge, their strength is far exceeded by the non-Keplerian potential of the binary (and similarly for the effect of pressure gradients in the disc). As the only stable orbits in this potential are precessing, the gas orbits do so too and their motion is well approximated by that expected for a massless test particle.

\section{Discussion}\label{sec:discussion}

\subsection{Numerical limitations and omissions from the model}\label{discussion:limitations}

There are a number of considerations that must be made in interpreting our results, given the simple disc model and the moderate (and in places poor) resolution of our primary simulation. In the region of uniform precession ($R \sim 5\,a_{\mathrm{b}}$), we resolve the disc vertical structure into 4 midplane smoothing lengths, narrowly satisfying the requirements of \citet{nelson06} to avoid under-estimating the midplane density. At larger and smaller radii (most especially inside the cavity), we fall below this and so our vertical disc structure in these regions is likely incorrect. However, vertical resolution is not the primary problem within the cavity as the gas there no longer behaves as if it were in a disc. The poor vertical resolution at large radius equally does not affect our results, as the high-resolution re-run is resolved here and shows no differences.

\begin{figure}
\includegraphics[width=\columnwidth]{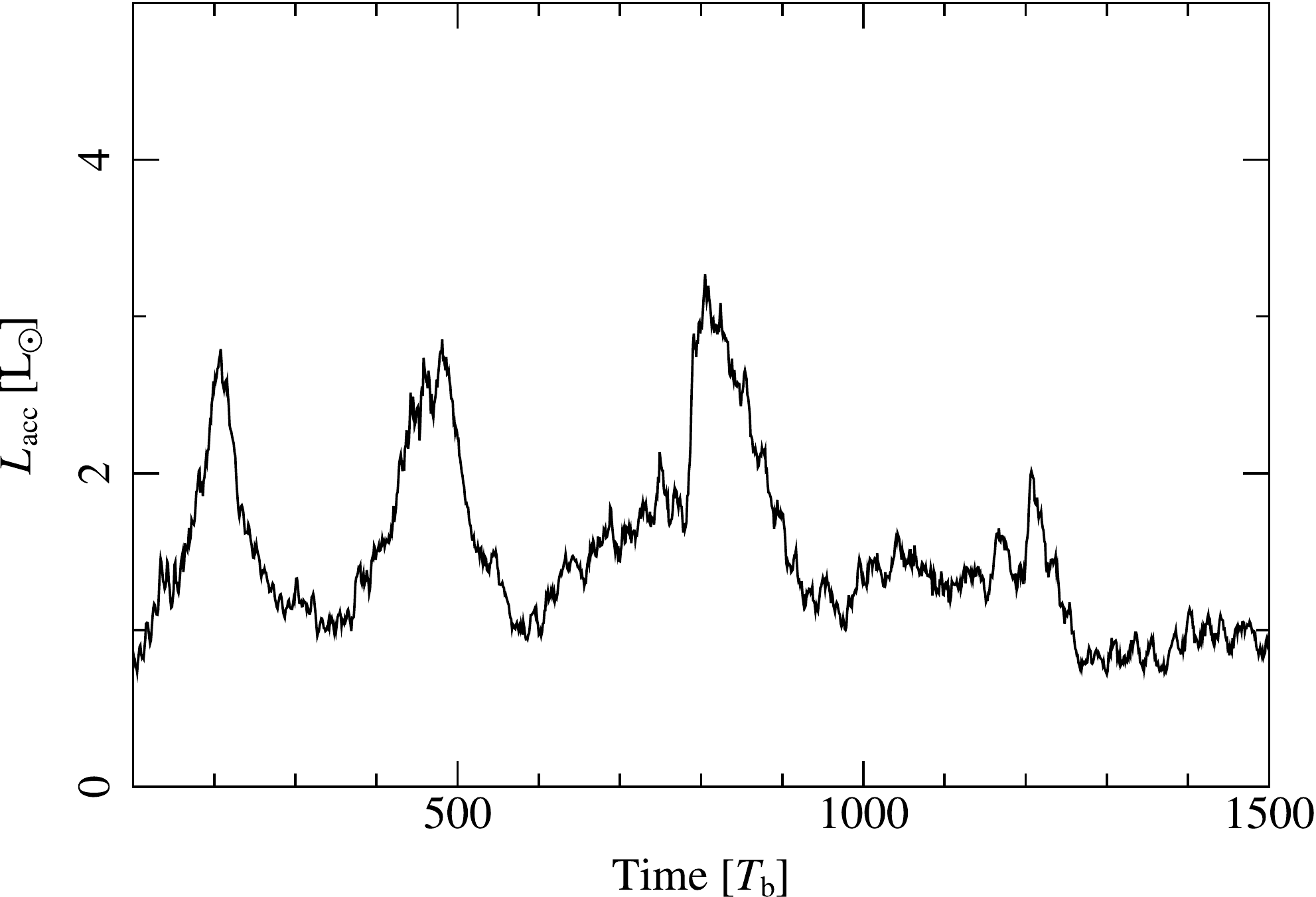}
\caption{Estimated luminosity evolution of the HD 104237 binary using Equation \ref{eq:lum}. We use values of $\dot{M}$ taken from Figure \ref{fig:7}, scaled down by a factor of $ 10^{3}$ to match the observed accretion rate of the system \citep{gradyetal04}. Values for the stellar masses were taken from \citet{garciaetal13} and stellar radii from \citet{fumelbohm12}. As the components cannot be distinguished at pericentre (where accretion peaks), we sum the luminosities of the components. The variability we find using this very simplified model is less than the uncertainty in the luminosity of the primary \citep[e.g.][]{garciaetal13}, so we would not expect the signal to be observable for the case of HD 104237.}
\label{fig:8}
\end{figure}

Of greatest concern is the poor resolution inside the cavity. In our main simulation with $N = 2 \times 10^6$ particles, each accretion stream typically consists of only a few thousand SPH particles -- as our code uses 50 nearest neighbours within the kernel, this is poor resolution indeed. In particular, in most of the stream the width is at best the same as the smoothing lengths of the particles in it, and wider than that towards the end of the stream being accreted.

The situation is less dire in the up-scaled run with higher resolution, where an accretion stream is typically $\gtrsim 10,000$ particles and is nearly always more than one smoothing length across. Given that Figures \ref{fig:5} and \ref{fig:7} show that both the disc and accretion follow the same pattern in the low- and high-resolution runs, the poor resolution within the cavity does not seem affect the results significantly. Indeed, the only difference is that the accretion rate is lower by approximately an order of magnitude, and this is a result of the better resolved accretion process and lower individual particle mass in the high-resolution case.

Although the accretion rate is lowered by increasing resolution, the pattern followed by the relative accretion rates onto the componants (i.e. $\dot{M}_{\mathrm{A}}/\dot{M}_{\mathrm{B}}$) remains unchanged. In calculating the potential change in accretion luminosity in Section \ref{sec:results}, we scale the raw accretion rates from our simulation to match the observed accretion rate seen in HD 104237. Although these raw numbers are unconverged between our simulations, it is instead this pattern in the relative accretion rates that dictates the signal shown in Figure \ref{fig:8}. Therefore it is unlikely that this signal is negatively affected by the unresolved accretion in our primary simulation.

Beyond these resolution issues, our assumption of a locally isothermal equation of state is problematic. This assumption may be valid far from the central binary, where the propagation time for photons scattering through the optically thick disc is longer than the dynamical time of the binary, but closer to it creates issues. In particular, \citet{marzarietal12} showed that the inclusion of a radiative treatment reduces the disc eccentricity, and has something of an effect on the disc's argument of periapse, though they did not run their simulations for long enough to investigate the precession effect we find. This reduced eccentricity could prevent the accretion variability pattern seen in our simulations from occurring, as the process relies on the pericentre of the cavity being significantly closer than its apocentre.

We also neglect proper consideration of how the accretion process itself proceeds. Correct thermodynamic treatment should consider where the gas shocks and how efficiently it is able to radiate this away, and as a result this can affect which binary component the gas is accreted onto \citep[e.g.][]{clarke12,youngetal14}. We also make no checks for boundness before swallowing gas particles, and both sink particles have the same accretion radius. Our neglect of good thermodynamic treatment means that particles being accreted do not have correct energies in any case, so testing if they are bound to the sink particle would be meaningless.

The size of any subdiscs around the stars is expected to be very small indeed, due to tidal interactions. A disc around the primary would be truncated at $R_{\mathrm{trunc}} = 4\,\mathrm{R}_{\sun}$, and the secondary disc at $R_{\mathrm{trunc}} = 3\,\mathrm{R}_{\sun}$ \citep{pichardoetal05,garciaetal13}. Comparing this with the stellar radii themselves \citep[respectively $3.3$ and $2.5\,\mathrm{R}_{\sun}$;][]{bohmetal04,fumelbohm12} clearly shows that any discs will be small indeed. Our neglect of them may well be justified then. If the stars have similar corotation radii, as suggested by \citet{garciaetal13}, then gas can be accreted directly onto the stars from this radius and so similar sink radii would be justified -- however it is unknown if this is the case. It is also not known what level of magnetospheric interaction occurs between the stars, and this could strongly influence how the accretion proceeds.

\subsection{Interpretation and observability}\label{discussion:interpretation}

We have found that circumbinary discs precessing around a strongly asymmetric binary will undergo changing accretion patterns on time-scales tied to the precession. In the case of HD 104237, the precession period is approximately 40 years, and so we expect that if it is observable, these changes will be seen on a time-scale about half this, around 20 years. In Figure \ref{fig:8} we made a very simple model of how the accretion luminosity at pericentre will evolve due to this precession effect.

The disc precession effect will be testable within a few years, given the high rate found in our simulations. In 2 years the disc will have precessed by approximately $18^{\circ}$, and by $36^{\circ}$ in 4 years, which should be easily detectable with \textit{VLTI} interferometry. The strong decentering of the cavity with respect to the binary should also be detectable. It may also be possible to determine the current state of the system by monitoring the Br$\gamma$ emission of each component over one binary period to determine which component is currently accreting at a higher rate -- however, this would require consistent monitoring over decades to give a high level of certainty about where in the precession-accretion cycle HD 104237 currently sits. A more detailed comparison of predictions of these simulations in the context of the HD104237 system will be conducted in a forthcoming paper (Dougados et al, in prep).

It is possible for this mechanism to play a key role in setting the mass ratio for eccentric binaries. Instead of primarily accreting onto the secondary, gas is able to accrete equally onto both components and maintain the mass ratio status-quo rather than driving the mass ratio towards unity, as is often seen in simulations of binary accretion \citep[e.g.][]{artymowicz83,bate00,devalborroetal11,bate14}. Indeed, in our simulation we find that the ratio of precession-period averaged accretion rates between the primary and secondary has values between $0.9$ and $1.3$, showing that in fact the primary can in fact accrete a little more gas using this mechanism than the secondary can.

\citet{bate14} found that with the inclusion of opacity effects in simulations of molecular cloud fragmentation it is possible to reproduce the observed mass ratio distribution well. Importantly, in these simulations there is a bias towards equal mass ratios in low eccentricity systems, and higher eccentricities correlate with more unequal masses. This is attributed in that paper to dissipation caused by circumbinary discs damping the eccentricity of equal mass systems, and given the resolution constraints of such large simulations this seems likely to be the case.

More noteworthy for our result is the work of \citet{halbwachsetal03}, who found in a survey of solar type (F-K) spectroscopic binaries that high mass ratio systems ($q_{\mathrm{b}} \geq 0.8$) have predominantly lower eccentricities than binaries of unequal masses. This is tentative evidence that the components of eccentric binaries accrete at comparable rates over long time-scales, as suggested by our simulations, allowing them to preserve their non-unity mass ratios. Further work is required in both quantifying how widely this effect operates in terms of binary mass ratio and eccentricity, and how significant the eccentricity differences are between observed equal- and unequal mass binary systems.

This phenomenon has applications for SMBH binaries in galactic centres, and indeed to date such circumbinary precession has only been explored in this context \citep[e.g.][]{macfadyenmilosavljevic08,shietal12,dorazioetal13,farrisetal14}. However, given that the time-scales in these systems are of the order of tens to hundreds of Myrs, it clearly is not going to be directly observable. It is still possible though that this mechanism helps maintain unequal masses for such binaries in the same manner as for stellar binaries. This is especially relevant as we expect these binaries to grow large eccentricities driven by disc interactions \citep[$e \sim 0.6$;][]{roedigetal11}. Therefore it is worth considering this effect when modelling observable signatures of candidate SMBH binaries rather than assuming that the majority of the accretion is onto the secondary as is standard \citep[e.g.][]{tanaka13}.

\section{Conclusions}\label{sec:conclusions}

We have presented the results of SPH simulations of the circumbinary accretion disc of the Herbig Ae binary star HD 104237 (DX Cha). We have found that the eccentric, unequal-mass binary causes the disc to become significantly eccentric at the inner edge and precesses uniformly with a rate $\dot{\varpi} = 0.48^{\circ}T_{\mathrm{b}}^{-1}$, giving a precession period of approximately 40 years. This eccentric shape to the cavity is consistent with observations that the cavity is not centred on the binary centre of mass, and the accretion pattern on single-orbit time-scales is also consistent with the observational evidence \citep{garciaetal13}. We find that this accretion pattern changes with the binary precession, as the position angle of the cavity with respect to the binary orbit changes which binary component accretes at the highest rate.

We model the changing accretion signature onto the binary as an accretion luminosity and find that, if the signal is observable, it should be seen on time-scales of 20 years (half the precession period). Although the change in luminosity is less than the uncertainty in current measurements of the stellar luminosities, if these uncertainties are systematic rather than intrinsic then the signal should still be observable.

We predict that this process will occur whenever a binary of sufficient eccentricity and extreme mass ratio is accompanied by an accretion disc. We find excellent agreement between the precession rates in our simulations and those predicted by \citet{leunglee13}, showing that their prescription can be used to predict the long-term accretion variability timescale for any such system. While we are unable to provide limits on what values for these parameters are necessary for the process to occur, we suggest that the observed trend towards increased eccentricity for unequal-mass binaries is evidence of this process at work \citep{halbwachsetal03}.

\section*{Acknowledgments}

We have used {\sc splash} \citep{price07} for the SPH visualization in Figures \ref{fig:1} to \ref{fig:4}. We would like to thank the anonymous referee for very helpful comments that helped to improve the paper. We acknowledge support from CONICYT-Chile through ALMA-CONICYT (311200007), FONDECYT (1141175), Basal (PFB0609) and Anillo (ACT1101) grants and from the Millennium Science Initiative (Chilean Ministry of Economy), through grant `Nucleus RC130007'. ACD also acknowledges support from the Science \& Technology Facilities Council (STFC) in the form of a PhD studentship.

The calculations were run on the DiRAC Complexity system, operated by the University of Leicester IT Services, which forms part of the STFC DiRAC HPC Facility (\href{http://www.dirac.ac.uk/}{www.dirac.ac.uk}). This equipment is funded by BIS National E-Infrastructure capital grant ST/K000373/1 and STFC DiRAC Operations grant ST/K0003259/1. DiRAC is part of the UK National E-Infrastructure.

\bibliography{mnrasmnemonic,references}
\bibliographystyle{mnras}

\label{lastpage}

\end{document}